\documentstyle[12pt,epsf]{article}

\newcommand{\eps}{\epsilon}
\newcommand{\ds}{\displaystyle}
\newcommand{\ra}{\rightarrow}
\newcommand{\be}{\begin{equation}}
\newcommand{\ee}{\end{equation}}
\newcommand{\bea}{\begin{eqnarray}}
\newcommand{\eea}{\end{eqnarray}}
\newcommand{\ci}{\cite}
\newcommand{\bi}{\bibitem}
\newcommand{\nono}{\nonumber \\}

\newcommand{\dd}{\partial}

\newcommand{\bfnabla}{\mbox{\boldmath$\nabla$}}

\newcommand{\vv}{\vec{\bf{v}}}
\newcommand{\uu}{\vec{\bf{u}}}

\def\dal{\,\lower0.3ex\vbox{\hrule\hbox{\vrule\kern2pt\vbox{\kern4pt\kern4pt}
\kern2pt\vrule}\hrule}\,}

\begin{document}

\title{\sl A differential equation for the Saffman-Taylor finger}
\vspace{1 true cm}
\author{G. K\"albermann\\Soil and Water department, Faculty of
Agriculture, Rehovot 76100, Israel}
\date{}
\maketitle
\begin{abstract}

{\noindent}We develop a stream function approach for the horizontal
Hele-Shaw, Saffman-Taylor finger.
The model yields a nonlinear time-dependent differential equation.
The finger widths derived from the equation 
are $1>\lambda>\frac{1}{\sqrt{5}}$, 
in units of half the width of the Hele-Shaw cell, in accordance
with observation. 
The equation contains the correct dispersion relation for the creation
of the finger instability.
In an accompanying paper the stationary solutions of the equation 
are found numerically.

\end{abstract}

\newpage
\section{\label{intro} Introduction}
The Hele-Shaw cell experiment of a less viscous fluid
displacing a more viscous one, is the paradigm of fingering phenomena\ci{pel}.
Saffman and Taylor\ci{st} found fingers that arise in the Hele-Shaw cell when 
oil or air or water, penetrate into oil or glycerin. An initial instability
develops into a finger or several competing fingers.
They compared the finger
profiles to an expression derived from the analytical properties of the
complex fluid potential and the stream function.
The agreement with the data was reasonable for large capillary numbers.

Since then, a large body of works has added to our knowledge
of the fingering phenomenon in various branches of the dynamics of continuous 
media, such as dendritic growth, directional solidification, diffusion-
initiated aggregation, flame front propagation, electromigration, as well
as fingering in porous media.
 
The flow in porous media prompted the initial research of fingering phenomena.
The topic is of the utmost importance in problems of
transport in saturated and unsaturated soils, groundwater pollution, etc.
\footnote{The website http://www.maths.ox.ac.uk/~howison/Hele-Shaw
/helearticles.bib
, cited in ref.\ci{ho20} carries an extensive (more than 600 papers), 
up to date list of references on the Hele-Shaw problem.}

Despite considerable efforts, the fingering phenomenon remains 
in many aspects uncharted territory and it defies intuition.\ci{tan}
As Tanveer\ci{tan} points out, small
effects, like local inhomogeneities, thin film effects, and 
surface tension make the theoretical description very difficult. 

The original theoretical formulation of Saffman and Taylor\ci{st} was ill-posed 
mathematically.\ci{ho86}
This is reflected in the indetermination for the asymptotic size of the finger. 
Contrarily to the measured profiles, that
were found to be bounded from below, by a size of 
around one half of the cell
width, the analytical expressions showed no such lower bound.

McLean and Saffman\ci{mc} improved the theoretical approach of
Saffman and Taylor\ci{st} by including the effect of
surface tension.
Tanveer\ci{tan} has shown that the approach of McLean and Saffman\ci{mc}, 
is equivalent to an expansion
in a parameter related to the finger half-width, 
capillary number and aspect ratio. 
This technique fails at the tail of the finger and higher order 
terms are needed. The results of McLean and Saffman\ci{mc} 
predicted profiles that matched very well the front (nose) part of the finger.
They found a limit of $\lambda>\frac{1}{2}$ to the finger half-width.
More recent experiments with different aspect ratios of the Hele-Shaw cell 
found the finger width limit is more likely around 
$\lambda >0.45$ of the cell width\ci{tab}.

There appear to be, other branches of solutions (two at least)\ci{van},
that compete with the branch found by McLean and Saffman\ci{mc}.
These branches are unstable\ci{kes,ben,tan1}.
Pitts\ci{pit} took advantage of the observed dependence of the curvature
of the finger on the angle, to obtain an analytical scale covariant
expression for the finger shape that fitted measured values (by Saffman
and Taylor and by Pitts himself) extremely well, especially for finger
half-widths smaller than $\lambda\approx 0.8$. For wider fingers, 
the profile function
was found to miss the measured finger by a small amount.
Pitts presented analytical expressions 
for the finger profile including the effect of surface tension.
He also derived  an expression for the
finger's asymptotic half-width containing both the capillary number, and
a fitted parameter. This parameter takes into account the
fluid films left behind by the passing finger.
Despite Pitts's success in reproducing the data, 
he did not provide a physical basis for the phenomenological assumptions.

DeGregoria and Schwartz\ci{deg} used a boundary integral Cauchy technique to 
investigate the production and propagation of fingers in the
Hele-Shaw cell.
Tryggvason and Aref\ci{trig} used a {\sl vortex in line}
method to determine the relationship between 
finger half-width and flow parameters. Both works, as well as the stationary
calculation of McLean and Saffman\ci{mc} give similar results
concerning the finger half-width dependence on capillary number.
A marked improvement is found in the numerical calculations
of Reinelt\ci{rei}.
  
The present work offers an alternative path to the
fingering problem.
We discuss the validity of Darcian flow in the Hele-Shaw cell. 
We find that the Hele-Shaw Darcian equation is valid 
in the direction of motion of the finger exclusively.
Consequently, there is no harmonic condition on the hydrodynamic potentials
that would prevent the existence of vortices at the finger
boundary.
 
From minimal assumptions and a convergent expansion of the stream function, 
we derive a nonlinear differential equation 
for the stationary finger. The equation is then generalized
to the time dependent case. We connect to the 
nonlinear hydrodynamic equation of Korteweg and deVries\ci{kdv}.
Numerical profiles for the stationary
fingers and finger properties are displayed in the following paper.

Section 2 discusses our departure from the conventional approach to the
Hele-Shaw, Saffman-Taylor theory.
Section 3 develops the stream function at the interface.
Time independent and time dependent
nonlinear equations for the finger, are obtained in section 4.
Section 5 states our conclusions.

\section{\label{streamf}Fluid Equations}

The dynamical creeping flow equation for the Hele-Shaw cell is
\ci{pel,tan}.

\be\label{creep}
\vv=-\frac{b^2}{12~\mu}\vec{\bfnabla} p
\ee

{\noindent}where $\vv$ is the viscous fluid velocity in the plane of the cell,
{\sl b} is the thickness of the cell, $\mu$ the viscosity, and {\sl p},
the pressure in the fluid.
Eq.(\ref{creep}) is obtained by averaging the Navier-Stokes
equations over the smallest dimension of the Hele-Shaw cell.
However, the true and real problem is inherently
three-dimensional\ci{trig}, especially because variable width films of fluid are
left behind the advancing finger.
The finger thickness is not constant comparing tail and nose. It is thinner
at the nose and thicker at the tail. 
Reinelt\ci{rei} dealt with these 
films of fluid by means of a three (or four) domain splitting of the Hele-Shaw
cell. This method improved the agreement between theory and experiment.

Eq.(\ref{creep}) imitates Darcy's
law for the flow in porous media\ci{st}.
Both displacing and displaced fluids are supposed to obey a Darcian law.
Eq.(\ref{creep}) implies a vanishing vorticity and consequently a harmonic
equation for either the velocity potential or the stream function.
As noted previously in the literature \ci{trig}, the neglect of vorticity
is unjustified at the front.

We here summarize the derivation of eq.(\ref{creep}) and conclude that
it is mainly valid for the longitudinal direction of the cell.

The Navier-Stokes equation in a horizontal cell
for a Newtonian fluid\ci{batch} without gravity is given by

\bea\label{Navier}
\rho\frac{D\uu}{Dt}=-\bfnabla p+\mu\bfnabla^2\uu
\eea

{\noindent}where 
$\ds\frac{D\uu}{Dt}~=~\frac{\dd~\uu}{\dd t}+\uu\cdot\bfnabla(\uu)$ 
denotes the material derivative, $p$ is the pressure
and, $\mu$ the viscosity coefficient.
For the stationary case, there is no explicit
time dependence. Also, higher order quadratic terms
in the velocity are neglected, in
 the creeping flow Stokes approximation.
The left hand side of eq.(\ref{Navier}) then equals zero.

Eq.(\ref{Navier}) becomes

\bea\label{nav1}
\bfnabla p=\mu\bfnabla^2\uu
\eea

The Hele-Shaw cell is usually a rectangular paralellepiped with a long 
dimension in
the direction of the flow, the {\it x} axis , a much smaller width,
 the {\it y} axis, and the smallest dimension by far, the thickness
of the cell, the {\it z} axis.
The aspect ratio
of the cell, the ratio of the width along the {\sl y} axis
as compared to the thickness along the {\sl z axis}
is large. The cell is designed to be very thin. 
The most important contribution to the Laplacian in eq.(\ref{nav1})
is the second derivative with respect to this small dimension, the
z axis, and to a lesser undetermined extent, the $\ds \frac{\dd^2}{\dd y^2}$
 term. The Poisseuille 
parabolic profile flow with separable dependence of the velocities
of the form (Eq.(4.8.19) in \ci{batch})

\bea\label{pois}
\uu&=&-\frac{1}{2\mu}\bfnabla p(x,y)~z~(b-z)
\eea

{\noindent}solves eq.(\ref{nav1}) with complete neglect of the 
derivatives with respect to {\sl x and y}.
Eq.(\ref{creep}) is then obtained by {\sl averaging} 
eq.(\ref{pois}) with respect to the
transverse dimension {\sl z}. This brings in a factor of 6
in the denominator and a factor of 12 in the final equation (\ref{creep}).

For eq.(\ref{pois}) to be valid, the pressure
gradient has to extend to infinity\ci{batch}. 
This condition holds for the flow {\it x} direction.
However, it is not appropriate for the transverse direction. 
In the {\sl y} direction there appear strictures comparable to the cell 
thickness between the finger and the lateral walls.
Pressure gradients in the {\it y} direction can be quite important. Even
inertia effects may become relevant for the {\it y} component of
the fluid equation of motion.
The averaging procedure, that lead to eq.(\ref{creep}) is also known
as the lubrication approximation.\ci{batch}
In the lubrication approximation, even 
in the context of thin films\ci{oron}, it is 
used for the direction of the flow only.
The {\it x} axis is ilimited, whereas the {\it y} axis has a definite
scale. That scale should be present in a correct treatment as a separate
length scale, besides the thickness of the cell {\it b}. However, and due
to the averaging procedure, it does not. These arguments suggest that
the transverse component of eq.(\ref{creep}) is not correct.

In the following, we will use eq.(\ref{creep}) for the direction
of motion of the finger only. 
The flow in the transverse direction will be constrained
by means of boundary conditions. 

The Darcian creeping flow equation we adhere to is

\bea\label{creep1}
u(x,y,t)=-\frac{b^2}{12~\mu}\frac{\dd~p}{\dd~x}
\eea

{\noindent}where {\it u} is the velocity in the {\it x} direction, and {\it p}
is the pressure averaged over the {\it z} coordinate.
The equation for the velocity in the {\it y} direction will be left in the
Navier-Stokes form (\ref{nav1}).
Contrarily to eq.(\ref{creep}), Eq.(\ref{creep1}), does allow vorticity
on the interface.

Both eqs.(\ref{creep},\ref{creep1}) are not free of inconsistencies.\ci{deg}
At the lateral solid edge of the cell,
$\ds y=\pm\frac{w}{2}$, with {\it w}, the width of the cell, we should demand
impenetrability of the fluid and no-slip, namely

\bea\label{noslip}
\frac{\dd~p}{\dd x}&=&0\nono
\frac{\dd~p}{\dd y}&=&0
\eea

Integrating the first of eq.(\ref{noslip}) form the entrance
of the channel to its exit, we find that the pressure is identical
at both ends on the lateral side of the cell. The contradiction arises
form the no-slip boundary condition applied to the creeping
flow equation (\ref{creep1}). This conundrum is usually resolved by ignoring
the no-slip condition without further justification.\ci{deg}
Although, the use of no-slip boundary condition has been contested repeatedly,  
because of the non-integrable stress\ci{huh,dussa,hock} it generates,
\footnote{A possible theoretical 
resolution of the no-slip problem for a moving contact line 
may be found in ref.\ci{pomeau}}
and recent experimental work on the subject\ci{granick,hervet} supports
this view, it is not consistent to apply it for the
the {\it z} direction and not for the {\it y} direction. 
Both directions are limited by the solid boundary of the cell.
In the next section we propose a method for the  partial restitution of
the no-slip condition.

Eq.(\ref{creep1}) is supplemented by boundary conditions.

At the interface, we have\ci{rei,park,tan}

\be\label{cond1}
\Delta p=\frac{2~T}{b}~cos\gamma+\frac{T}{R}
\ee
 
{\noindent}
where, {\sl T} is the surface tension parameter of the Young-Laplace formula,
{\sl R}, the radius of curvature in the plane of the cell and $\gamma$, the
contact angle between the finger and the cell wall in the transverse {\sl b}
direction. 
The first term in eq.(\ref{cond1}) is usually dropped, because the
flow depends on the gradient of the pressure.
The term should be kept for three dimensional treatments that consider 
the variation of contact angle $\gamma$ and, the change of finger
thickness as one proceeds from nose to tail.
Pitts\ci{pit} circumvented the problem by introducing an effective parameter 
to account for the difference in pressure inside the fluid between tail
and nose.
The narrower the finger as compared to the
channel width, the stronger the effect of films of fluid left behind\ci{rei}.
In our minimal model we drop the first term in eq.(\ref{cond1}).
Therefore, our results are better suited for wide fingers and 
small capillary numbers.

Equation (\ref{cond1}) reduces then to

\be\label{cond11}
\Delta p=\frac{\tilde T}{R}
\ee

{\noindent}with ${\tilde T}$ an effective surface tension parameter.

To complete the set of equations we assume the fluids to be immiscible.
The velocity of the fluid at the interface has to be tangential to the boundary.
As discussed in the next section, the immiscibility condition
implies that the boundary is a stream-line.

Finally, for an incompressible fluid of constant density,
the continuity equation reads

\bea\label{continu}
\vec{\bfnabla}\cdot\vv=0
\eea

The stream function $\ds \vec{\Psi}$, determines
the fluid flow through $\ds \vv~=\vec{\bfnabla}{\bf x}{\vec{\Psi}}$.
The stationary equation of continuity of eq.(\ref{continu}) is therefore 
obeyed by construction
 $\ds \vec{\bfnabla}\cdot\vec{\bfnabla}{\bf x}{\vec\Psi}~=~0$.

\section{\label{str1} The stream function}

The interface between the displaced and displacing fluids in the
Hele-Shaw cell carries vorticity\ci{trig}.
Consider figure 1, where a typical schematical profile of a finger
in a Hele-Shaw cell is depicted.
The tongue of less viscous fluid (a gas for instance) penetrates a more viscous
fluid to its right. The picture shows a cross section or lateral
view of a horizontal Hele-Shaw cell. The {\sl z} coordinate is along the thin
vertical direction. We have drawn a few vortices to guide the eye.
\begin{figure}
\epsffile{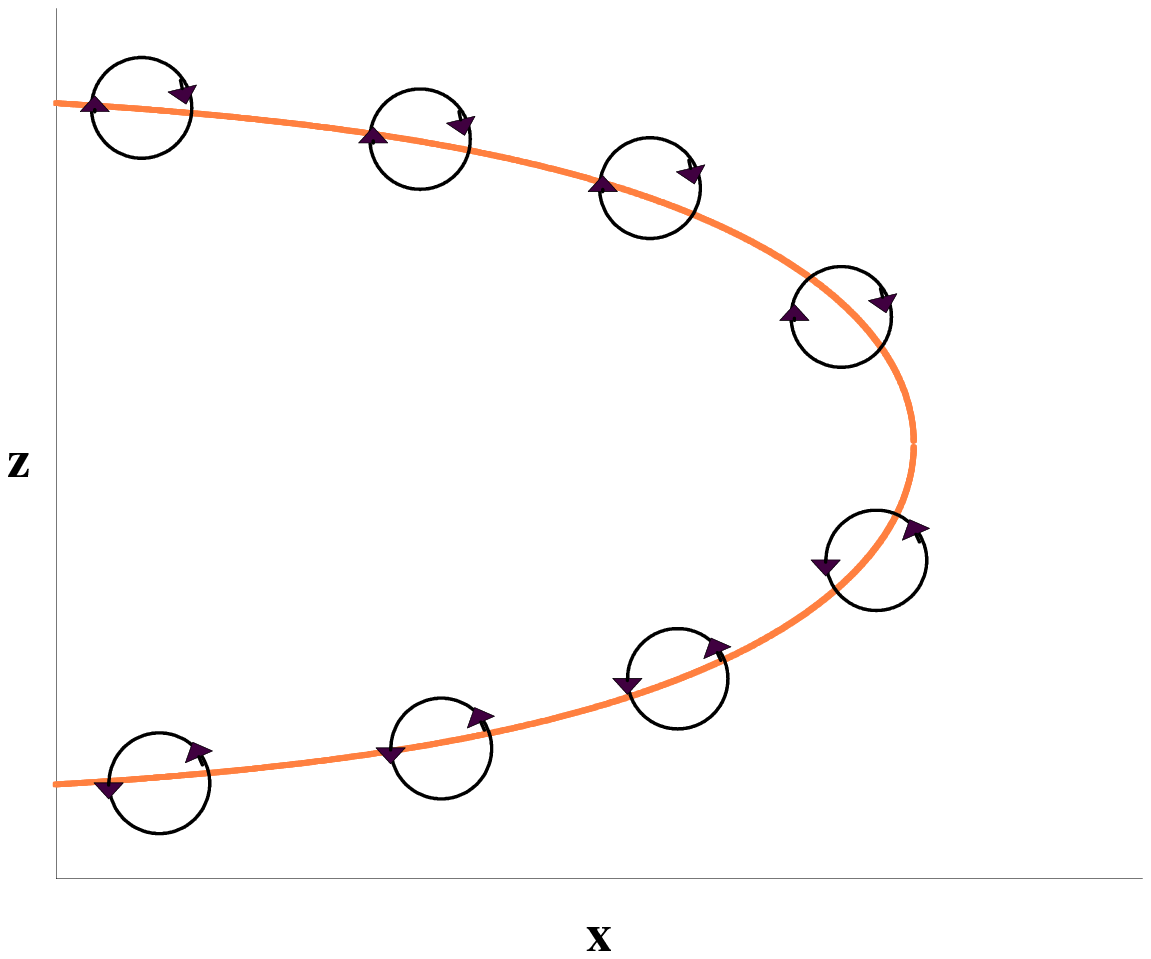}
\vsize=5 cm
\caption{\sl Vortices on a transverse slice of a finger interface.}
\label{fig1}
\end{figure}

Recent experimental works \ci{vere1,vere2}, have shown 
that finger flow involves vorticity. 
Gravity driven fingers advance by
a kind of rolling motion, as suggested long ago by Yarnold\ci{yarnold}, 
and West\ci{west} in the context of capillary rise.

The all important property of the fluid flow, is not the irrotationality, 
broken at the interface and elsewhere, 
but the solenoidality originating from the incompressibility condition. 
The stream function, is better suited for the theoretical analysis
of the problem.

We here determine the stream function at the interface
by means of a convergent perurbation expansion.
The stream function together with the Young-Laplace equation 
yield nonlinear differential equations
for stationary and time dependent fingers.

We first renormalize the axes to 
$\ds x\ra{\tilde x}=
\frac{x}{w/2},~y\ra{\tilde y}=\frac{y}{w/2}$
with $w$ being the width of the Hele-Shaw cell.
As a consequence of this rescaling, the finger
transverse dimension becomes less than one, 
and serves as an expansion parameter.

Two conditions constrain the stream field:
The impenetrability of the cell wall and the immiscibility of the fluids.
 
The corresponding stream function boundary conditions read

\bea\label{constraint1}
\frac{\dd\Psi}{\dd x}=0
\eea

{\noindent}at $\ds y=\pm 1$, and 

\bea\label{constraint2}
\frac{\dd\eta}{\dd t}+\frac{\dd\Psi}{\dd y}\frac{\dd\eta}{\dd x}=
-\frac{\dd\Psi}{\dd x}
\eea

{\noindent}at $y=\eta$, where $\eta(x,t)$ denotes the interface curve.
Eq.(\ref{constraint2}) may be easily derived by writing the 
differential of $\eta$ as $\ds 
d\eta=\frac{\dd\eta}{\dd t}~dt+\frac{\dd\eta}{\dd x}
~dx$ and using the definitions 

\bea\label{vels}
v_x&=&\frac{dx}{dt}~=~\frac{\dd\Psi}{\dd y}\nono
v_y&=&\frac{dy}{dt}~=~-\frac{\dd\Psi}{\dd x}
\eea

{\noindent}For the stationary finger, in its rest frame we have 
$\ds\frac{\dd\eta}{\dd t}~=~0$. 
With this substitution, eq.(\ref{constraint2}) becomes

\bea\label{const5}
\frac{\dd\Psi}{\dd y}\frac{\dd\eta}{\dd x}=-\frac{\dd\Psi}{\dd x},~at~y~=~\eta
\eea

The differential of $\Psi$ 
$\ds d\Psi=\frac{\dd\Psi}{\dd y}~dy+\frac{\dd\Psi}{\dd x}~dx$, 
together with eq.(\ref{const5}) determine
the unique solution for the stream function in the stationary
case at the interface to be 
\be\label{psi0}
\Psi_{stationary}~=~constant
\ee

Without loss of generality, this constant may be taken to be equal
to zero, because the stream function enters the calculations only
through its derivatives.

An alternative expression of eq.(\ref{const5}) is

\be\label{cond2}
\alpha(\vv)=tan^{-1}(\eta')
\ee

{\noindent}where $\alpha$ is the angle between the fluid velocity and 
the axis of propagation of
the finger (x axis) and $\eta'$ is the derivative to the finger (y axis)
with respect to the x axis.

For the time-dependent case, the solution of eq.(\ref{constraint2}) reads

\be\label{const1}
\Psi(x,y,t)=\Psi_0(x,y)-\int{dx~~\frac{\dd\eta}{\dd t}}
\ee

Eq.(\ref{const1}) solves eq.(\ref{constraint2}), with $\Psi_0$ given by 
the appropriate solution of eq.(\ref{const5}).

At the front, we can rewrite the eq.(\ref{const1}) in the form

\be\label{const12}
\Psi(t)=\Psi_0-\int{dy~\frac{\dd\eta}{\dd t}/\frac{\dd\eta}{\dd x}}
\ee

{\noindent}where $\eta(x,t)$ is understood as a function of {\sl x and t} only.
Eq.(\ref{const12}) shows that the value of $\ds \Psi$ at the interface
changes with time. This equation is needed for the development of the time
dependent finger equation of motion.

We now proceed to derive the stream function at the interface for
the stationary case corresponding to a fully developed finger 
traveling at constant
speed, to be generalized later to the nonstationary situation. 

Consider the stream function in the finger rest frame. The finger
is traveling in the positive {\sl x} direction at constant speed {\sl U}.
At long distances ahead of the finger the fluid is assumed to flow
with a constant velocity {\sl V}. As discussed in section 2, 
the no-slip condition cannot be
imposed in the Hele-Shaw cell and consequently the velocity {\sl V} is 
the same all over the cell width including at the lateral boundary.
This is a reasonable assumption provided the viscosity of the fluid is small.
However, for large capillary numbers, or large viscosity, this
procedure is unreasonable. Boundary layers next to surfaces are
patent in common phenomena even for moderately viscous fluids.

We follow  a more conservative approach and demand the
stream line next to the lateral edge to carry a velocity {\sl V} that
is {\sl not} the asymptotic velocity of the fluid ahead of the finger.
It depends on the flow properties.
For capillary numbers tending to infinity it has to be equal to zero.
Large capillary numbers may be implemented by
using very viscous fluids. For such fluids, the no-slip condition is a must.
Therefore in the limit of infinite capillary number {\sl V=0}.
In the opposite limit, i.e. no viscosity at all (finger half-width going to
the full cell width) it is equal to the finger velocity. Hence
we can assert $0<V<U$.\footnote{Reinelt's equations\ci{rei} carry
an extra term at the lateral edge that transforms the velocity
{\sl V} to a variable.}
The parameter {\sl V} of our slip boundary condition will be 
determined selfconsistently by the dynamics, it is {\sl not} a free parameter.

Expanding the stream function to $\ds O(y^5)$ with a separable ansatz, 
antisymmetric in {\sl y} for symmetric fingers, 
and in the finger rest frame, we find 
(we drop from now on the tilde on {\sl x,y}, i.e. we work with rescaled
coordinates)

\bea\label{stream}
\Psi(x,y,t)=y~(V(x,t)-U(x,t)+A(x,t)+B(x,t)~y^2+C(x,t)~y^4+...)
\eea

with {\sl A, B, C} unknown functions. 

Eq.(\ref{constraint1}), as well as the constancy of the fluid velocities, for
the stationary finger in its rest frame imply

\bea\label{dem}
v_y&=&-\frac{\dd\Psi}{\dd x}=0~,at~~y~=~\pm 1\nono
v_x&=&\frac{\dd\Psi}{\dd y}=~V-U~,at~~y~=~\pm 1
\eea

Inserting eq.(\ref{dem}) in eq.(\ref{stream}), we are able
to fix the functions {\sl B and C} in terms of {\sl A}.
Explicitly
\bea\label{sldem}
0&=&A+3~B~+5~C~\nono
0&=&~A'+B'+C'
\eea

primes denoting derivatives with respect to {\sl x}.

Recalling that constants are irrelevant for the stream function, 
the unique solution to this order becomes 

\bea\label{soldem}
B&=&-2~A\nono
C&=&A
\eea

The stream function of eq.(\ref{stream}) with the conditions of 
eqs.(\ref{sldem},\ref{soldem}) reads

\be\label{stream1}
\Psi(x,y)=(V-U)~y+A(x)~y~(1-y^2)^2+...
\ee

{\noindent}with

\bea\label{cond4}
\Psi=0~~at~y=\eta
\eea

In order to assess the convergence of the separable expansion, 
we consider the the stream function to $\ds O(y^{11})$. Such a high order
is needed for the fulfillment of the boundary conditions. Proceeding in
the same manner that lead to eq.(\ref{stream1}), we find

\bea\label{streamh}
\Psi_1(x,y,t)=y~\big[V(x,t)-U(x,t)+\big(A(x,t)~+y^6~D(x,t)\big)~(1-y^2)^2\big]
\eea

{\noindent} with $\ds D(x,t)$ an unknown function.
Only the {\sl x} component of the
velocity is needed in eq.(\ref{creep1}). Including the higher order
terms of eq.(\ref{streamh}) this velocity becomes

\bea\label{v1}
v_x=V(x,t)-U(x,t)+(1-y^2)~(A(x,t)~(1-5~y^2)+y^6~D(x,t)~(7-11~y^2))
\eea

However, for the stationary finger, we demanded $\ds \Psi=0$ at the boundary
(\ref{cond4}), then $\ds D(x,t)=0$ at the interface. 
Therefore, for the velocity at the interface in the longitudinal direction
eq.(\ref{stream1}) is exact.
The equivalence between eqs.(\ref{stream1},\ref{streamh}) 
breaks down for locations off the interface.
Nevertheless, even inside the fluid, the $\ds y^6$ dependence with $\ds y<1$
makes the contribution of the higher order term unlikely.
These terms correspond to higher order fluid velocity gradients. 
These gradients become relevant when there exist large stresses
inside the fluid. However, eq.(\ref{creep1})
was derived under the assumption of Poisseuille creeping flow, meaning
an orderly stationary almost constant very slow flow with negligible
internal stresses in the plane of the cell. The large stress
situation is beyond the scope of the problem at hand and higher
order terms should be ignored even inside the fluid.

In the next section we use eq.(\ref{creep1}) to derive a differential
equation for the stationary finger and another one for the time
dependent situation.

\section{\label{equ}Nonlinear differential equations for the finger}

Inserting eq.(\ref{cond4}) in  eq.(\ref{stream1}), 
$A(x)$ is determined at the front to be

\bea\label{ax}
A(x)=\frac{U-V}{{(1-\eta^2)}^2},~ at~~y=\eta(x)
\eea

Using eqs.(\ref{vels},\ref{stream1},\ref{ax}),  
the velocity of the fluid at the static finger, 
the only place where we can determine {\sl A} in closed form, becomes

\bea\label{vel}
v_{x,static}&=&(V-U)+A(x)~(1-\eta^2)~(1-5~\eta^2)\nono
v_{y,static}&=&-\frac{\dd A(x)}{\dd x}~\eta~{(1-\eta^2)}^2\nono
A(x)&=&\frac{U-V}{{(1-\eta^2)}^2}
\eea

{\noindent}with $\eta$, a function of x.

The rescaled curvature in eq.(\ref{cond11}) is 

\bea\label{curv}
\frac{1}{R}=\frac{\eta''}{(1+\eta'^2)^{\frac{3}{2}}}
\eea

Where primes denote derivatives with respect to x. 

The sign of the curvature is the appropriate one.
This can be seen by using eq.(\ref{creep1}). The left hand side of
the equation is a the positive velocity of the finger, chosen
here to move from left to right, therefore
the curvature has to decrease along {\sl x}. In the upper part
of the finger the curvature is negative and becomes more so
as we proceed along {\sl x}. Therefore the right hand side
of eq.(\ref{creep1}) is positive with the positive sign in eq.(\ref{curv}).

Eq.(\ref{curv}) is the standard expression 
for the curvature, and, it is easily derivable from the definition
of the arclength and the corresponding angle for differential
increments. 

Equations (\ref{creep1},\ref{cond11},\ref{curv}) after rescaling imply

\bea\label{diff0}
v_x=-~\frac{\tilde T~b^2}{3~w^2\mu}\frac{\dd}{\dd x}
\bigg[\frac{\eta''}{(1+\eta'^2)^{\frac{3}{2}}}\bigg]
\eea

The stationary finger equation is obtained now by
evaluating the derivative in eq.(\ref{diff0}), with 

\be\label{vx}
v_{x,static}=(V-U)~\frac{4~\eta^2}{1-\eta^2}
\ee

Transforming back to the rest frame
of the cell, with the finger in motion, $\ds v_x~=~v_{x,static}+U$.
we find

\bea\label{diff}
0&=&\eta_{xxx}-3~\frac{\eta_{xx}
^2~\eta_x}{1+\eta_x^2}~+~(1+\eta_x^2)^{\frac{3}{2}}~\frac{W(\eta)}{4~B}\nono
\eea
{\noindent}with
\bea\label{w}
W(\eta)&=&\frac{4~\eps~\eta^2+1-5~\eta^2}{1-\eta^2}\nono
\frac{1}{B}&=&\frac{12\mu~U~w^2}{{\tilde T}~b^2}\nono
\eps&=&\frac{V}{U}\nono
\eea
{\noindent}and the suffix in eq.(\ref{diff}) indicating 
differentiation with respect to {\sl x}.

An alternative expression in terms of the arclength measured 
from the tail of the finger reads

\bea\label{diffs}
0&=&\eta_{sss}+\frac{\eta_{ss}
^2~\eta_s}{1-\eta_s^2}+(1-\eta_s^2)~\frac{W(\eta)}{4~B}\nono
\eea

Customarily $\eps$ of eq.(\ref{w}) is equated to 
$\lambda$, the finger half-width. 
In section (\ref{str1}) we discussed the no-slip condition in the
Hele-Shaw cell and concluded that this procedure is inconsistent.
In the present work we leave $\ds \eps=\frac{V}{U}$ as a free
parameter to be determined self-consistently from the solutions to
the nonlinear equations.  Asymptotically far back at the tail
we must have $W(\eta)~=~0$, or, 
$\ds \epsilon=\frac{5~\lambda^2-1}{4~\lambda^2}$. $\ds \lambda$ determines
$\ds \epsilon$.

Eq.(\ref{diff}) implies a lower bound on the finger thickness.
In the limit of infinite capillary number $\ds \epsilon\ra 0$ and
at the tail of the finger $W(\eta)=0$ yields

\bea\label{lower}
\lambda>\frac{1}{\sqrt{5}}
\eea

This straightforward and simple result coincides with
experimental observation and improves the lower bounds found in
the literature by much more laborious means.

The potential {\sl W} guides the propagation of the finger. 
Eq.(\ref{diff}) is of third order in the spatial derivatives. 
Third order differential equations for
interface propagation are well known in the literature. Some notorious
examples are: Landau and Levich\ci{land}, Bretherton\ci{bre}, 
and Park and Homsy\ci{park}.

The third order equation (\ref{diff}) can be transformed to a more manageable
second order one in terms of the angle tangent to the curve, 
with $\eta$ obtained by integration. For $ds$ starting at the tail, 
where $\ds\theta\approx\pi$, $\ds ds~cos(\theta)=~-dx$.

Equation (\ref{diffs}) now reads

\bea\label{diff2}
0&=&\frac{{\dd}^2\theta}{\dd s^2}-cos(\theta)~\frac{W(\eta)}{4~B}\nono
\eta&=&\lambda-\int{ds~sin(\theta)}
\eea

With $W(\eta)$ defined in eq.(\ref{w}).

We now proceed to derive the time dependent equation.
Using eq.(\ref{const12}), 
and recalling the definition of $v_x$ in terms of the stream function of
eq.(\ref{vels}), we now have

\bea\label{dyna}
v_x=-\frac{\eta_t}{\eta_x}~+~U~+~v_{x,static}
\eea

Rescaling as before the length coordinates $\eta$, and {\sl x} 
by $\ds \frac{w}{2}$,
and the time by $\ds t\ra\frac{t~U}{w/2}$, and inserting eq.(\ref{dyna}) in 
eq.(\ref{diff0}), we obtain the time-dependent differential equation

\bea\label{time}
0&=&\eta_{xxx}-3~\frac{\eta_{xx}
^2~\eta_x}{1+\eta_x^2}~+~(1+\eta_x^2)^{\frac{3}{2}}~
\frac{1}{4~B}\bigg[{W(\eta)-\frac{\eta_t}{\eta_x}}\bigg]
\eea

In terms of the arclength eq.(\ref{time}) becomes

\bea\label{times}
0&=&\eta_{sss}+\frac{\eta_{ss}
^2~\eta_s}{1-\eta_s^2}+(1-\eta_s^2)~\frac{1}{4~B}\bigg[
W(\eta)-\sqrt{1-\eta_s^2}\frac{\eta_t}{\eta_s}\bigg]
\eea

While for the angle tangent to the front, the 
time-dependent equation (\ref{time}) reads

\bea\label{time2}
0&=&\frac{{\dd}^2\theta}{\dd s^2}-cos(\theta)~\frac{1}{4~B}
\bigg[W(\eta)+cos(\theta)~{\frac{\dd\theta}{\dd t}}/
{\frac{\dd\theta}{\dd s}}\bigg]\nono
\eta&=&\lambda-\int{ds~sin(\theta)}
\eea

The time-dependent equation (\ref{time}) gives the correct
dispersion relation for the perturbation of a planar front
as we now show.

The dispersion relation we seek is\ci{chuoke}
$\ds\sigma=|\alpha|~(1-\lambda\alpha^2)$
with $\ds \sigma$ the instability parameter, $\alpha$ related to
the wavenumber of the perturbation and $\lambda$ a parameter
 proportional to the surface tension.
The velocity, that is proportional to $\ds \alpha$,
destabilizes and the surface tension stabilizes the perturbation.

Consider eq.(\ref{time}) at $y\approx 0$ for a perturbation protruding 
from a moving front 

\bea\label{timep}
0&=&\eta_{xxx}-3~\frac{\eta_{xx}
^2~\eta_x}{1+\eta_x^2}~+~(1+\eta_x^2)^{\frac{3}{2}}~
\frac{1}{4~B}\bigg(1-\frac{\eta_t}{\eta_x}\bigg)
\eea

Eq.(\ref{timep}) can be rewritten as

\bea\label{timepp}
\frac{\dd}{\dd x}~\chi&=&0\nono
\chi&=&4~B\frac{\eta_{xx}}
{(1+\eta_x^2)^{\frac{3}{2}}}~+x~-\int{dx~\frac{\eta_t}{\eta_x}}
\eea

Following Chuoke et al.\ci{chuoke} we consider the perturbation

\bea\label{ansatz}
x(y,t)&=&\beta(\alpha)~e^{\phi}\nono
\phi&=&\alpha~y+\sigma~t
\eea

{\noindent}$\beta$ is the amplitude of the perturbation, a function of
$\alpha$\ci{chuoke}.
The integral in eq.(\ref{timepp}) may be rewritten as

\bea\label{int1}
I=\int{dx \frac{\eta_t}{\eta_x}}~=~\int{x_y~x_t~dy}~~~at~~~y=\eta
\eea
Evaluating the integral (\ref{int1}) with eq.(\ref{ansatz}) we find

\bea\label{int2}
I&=&\frac{\sigma}{2}~\beta^2~e^{2\sigma~t}~(e^{2\alpha y}-1)\nono
&\approx&\sigma~\beta^2\alpha~y
\eea

The curvature may be calculated now by using 

\bea\label{curv2}
\frac{\eta_{xx}}{(1+\eta_x^2)^{\frac{3}{2}}}
&=&-\frac{x_{yy}}{(1+x_y^2)^{\frac{3}{2}}}\nono
&\approx&-x_{yy}\nono
&=&-\alpha^2~x
\eea

Using eq.(\ref{curv2}), the integral of eq.(\ref{int2}), and,
 {\sl x} of eq.(\ref{ansatz}), at $t=0$ to $O(y^2)$, 
as we are assuming a perturbative expansion, $\chi$ of eq.(\ref{timepp}),
 becomes

\bea\label{chi0}
\chi_0\approx(-4~B~\alpha^2\beta~+~\beta)~(1+~~\alpha~y)-~\sigma~\beta^2
\alpha~y
\eea

The constant piece of $\chi_0$ in eq.(\ref{chi0}), $-4~B~
\alpha^2\beta~+~\beta$, is irrelevant. This is the reason we do not need
 to specify the lower bound of the integral in eq.(\ref{int1}). We took it
as $y~=~0$. Any other choice will merely change the unimportant constant in 
 $\ds \chi$.

To satisfy eq.(\ref{timepp}), to the lowest order in $y$, eq.(\ref{chi0}) 
has to obey the algebraic condition

\bea\label{chiob}
-4~B~\alpha^3\beta~+~\beta\alpha-\sigma\beta^2\alpha=0
\eea

The third term in eq.(\ref{chiob}) is quadratic in $\beta$. 
Independence from the initial perturbation amplitude, 
requires $\beta$ to be a function of $\alpha$, as assumed in eq.(\ref{ansatz}). 
We can now determine this functional dependence to be
$\ds \beta\alpha~=~k$, with {\sl k}, 
a constant. $\beta$ is a positive number, therefore,
{\sl k} has to be positive for positive $\alpha$, or
 negative for negative $\alpha$.

Equivalently, the proportionality above can be written as

\bea\label{beta}
\beta=\frac{|k|}{|\alpha|}.
\eea

Inserting eq.(\ref{beta}) into eq.(\ref{chiob}) we find

\bea\label{chiobp}
-4~B~\alpha^3\beta~+~\beta\alpha-\sigma\beta\alpha
\frac{|k|}{|\alpha|}=0
\eea

Finally, redefining $\ds |\alpha|\ra\frac{|\alpha|}{|k|},~
\lambda=4~B~k^2$ we find the expected dispersion relation

\bea\label{sigma}
\sigma=|\alpha|~(1-\lambda~\alpha^2)
\eea
{\noindent} the desired dispersion relation.

The time-dependent equation was developed on the basis of
an antisymmetric stream function, eq.(\ref{stream1}), the solutions, may, 
nevertheless, lack any predetermined symmetry. 
Although the equations are
symmetric under a certain transformation, the solutions can break the
symmetry.
Boundary conditions and initial conditions are key factors
in determining the symmetries of the solutions.
Such is the case of an initial 
asymmetric perturbation to the front.
On the experimental side, a symmetric finger develops
only asymptotically at long times. The transient behavior 
is asymmetric.

\section{Conclusions}

The contribution of the present work consists in
the development of nonlinear differential equations for the Hele-Shaw
Saffman-Taylor fingering problem.
The approach is based upon the treatment of the
stream function as a separable potential without any free 
adjustable parameters.

For the stationary case, the finger
nonlinear equations are eqs.(\ref{diff}), and (\ref{diffs}) 
the latter expressed in terms of the arclength,
while eq.(\ref{diff2}) is the corresponding integrodifferential equation, 
of second order, for the angle as a function of arclength. These are
nonlinear and nonlocal equations.

The time-dependent equation is of the third order in space, and 
first order in time.
Eqs.(\ref{time},\ref{times}), corresponds to $\eta$ in terms
of x and the arclength respectively, 
whereas eq.(\ref{time2}) is its integrodifferential
equation for the angle as a function of the arclength also.

The equations found here, resemble 
the Korteweg-deVries (KdV) equation, that possesses
solitary wave solutions\ci{kdv}.
The KdV equation describes
the nonlinear propagation of shallow water waves in a channel.

The KdV equation reads

\bea\label{kdv}
c_0\bigg(\frac{h^2}{6}-\frac{2~T~h}{\rho~g}\bigg)
\eta_{xxx}+\frac{3~c_0}{2~h}\eta\eta_x+\eta_x(c_0+\frac{\eta_t}{\eta_x})=0
\eea

{\noindent}where {\sl h}, 
is the depth of the channel, $c_0=\sqrt{g~h}$ is the speed 
of propagation of linear waves in the channel,
 {\sl g} is the acceleration of gravity and $\eta$ denotes the soliton
profile propagating in the channel along the {\sl x} axis with velocity 
$\ds U~=~-\frac{\eta_t}{\eta_x}$,
 in a fluid with density $\rho$ and surface tension {\sl T}.
This is a third order equation nonlinear in $\eta$. 
The specific nonlinear term, and,
the potential found in eq.(\ref{diff}), differ from the quadratic
nonlinear term of the KdV equation (\ref{kdv}).
The nonlinear terms in eq.(\ref{diff}), and eq.(\ref{kdv})
arise from the boundary conditions at the interface between the front 
and the displaced fluid as well as the surface tension pressure drop.
Eq.(\ref{diff}) however, carries also the information of the lateral
flow constraint that is absent in eq.(\ref{kdv}).
Moreover, eq.(\ref{diff}) is based upon the creeping flow assumption of eq.
(\ref{creep1}), while eq.(\ref{kdv}) does not assume such
a restricted version of the Navier-Stokes equation. Therefore
the nonlinear terms differ.

Nevertheless, the equations are similar. Both are nonlinear of third order
in the shape variable, and of first order in time.
Both equations 
describe the propagation of stationary and stable shape fronts in a 
fluid medium.

Links between soliton equations and the Hele-Shaw finger appear in 
a work of Kadanoff\ci{kad}, in which varieties of 
Harry-Dym equations were found to be related 
to finger development.
The above equations belong to a broad class
of {\sl flux-like} partial differential equations,
 widely used in the literature. In the context of Darcian
flow in the Hele-Shaw cell,
 Goldstein et al.\ci{gold} studied instabilities and singularities
by means of an equation of this type.
The main difference between the equations we derived here, and the one
of Goldstein et al.\ci{gold}, is, again, the appearance of
a potential term $W(\eta)$ in eqs.(\ref{time},\ref{time2}).
For the Hele-Shaw cell, 
the cell provides the potential for the propagation of the
finger, while the pressure difference determined by the surface tension
provides the dynamics.

\end{document}